\documentclass[superscriptaddress,showpacs,preprintnumbers,amsmath,amssymb,aps,twocolumn,prl,sort,compress,nofootinbib]{revtex4-2}

\usepackage[utf8]{inputenc}  
\usepackage[T1]{fontenc}
\usepackage{upgreek}

\usepackage{comment}
\usepackage{lipsum}
\usepackage{graphicx}
\usepackage{bm}
\usepackage{enumitem}
\usepackage{color}
\usepackage{multirow}
\usepackage{adjustbox}	
\usepackage{placeins}	
\usepackage{amsmath}
\usepackage{soul}

\usepackage[dvipsnames]{xcolor}
\usepackage[pdfpagelabels, pdfencoding=auto, psdextra]{hyperref}
 \hypersetup{%
  pdfsubject=Paper,
  pdfkeywords={nuclear physics} {Bayesian} {chiral EFT} {nuclear matter},
  unicode = true,
  breaklinks = true,
  colorlinks = true,
  linkcolor = blue,
  citecolor = blue,
  menucolor = blue,
  citecolor = blue,
  urlcolor = blue
 }

\usepackage{dcolumn}
\newcolumntype{d}[1]{D{.}{.}{#1}}   
\usepackage[modulo,columnwise,running]{lineno}

\newcommand{\tu}{\affiliation{\href{https://ror.org/042aqky30}{Technische Universität Dresden}, 01069 Dresden, Germany}}
\newcommand{\hzdr}{\affiliation{\href{https://ror.org/01zy2cs03}{Helmholtz-Zentrum Dresden-Rossendorf (HZDR)}, Bautzner Landstr. 400, 01328 Dresden, Germany}}

\newcommand{\sandiego}{\affiliation{Department of Physics, \href{https://ror.org/0264fdx42}{San Diego State University}, 5500 Campanile Dr, San Diego, CA 92182, USA}}

\newcommand{\michigan}{\affiliation{\href{https://ror.org/03r4g9w46}{Facility for Rare Isotope Beams}, \href{https://ror.org/05hs6h993}{Michigan State University}, East Lansing MI, 48824, USA}}

\newcommand{\INFN}{\affiliation{\href{https://ror.org/00z34yn88}{Istituto Nazionale di Fisica Nucleare, Sezione di Padova}, via Marzolo 8, 35131 Padova, Italy}}

\newcommand{\padova}{\affiliation{Dipartimento di Fisica dell'Universit\`a di Padova, via Marzolo 8, 35131 Padova, Italy.}}

\newcommand{\atomki}{\affiliation{HUN-REN Institute for Nuclear Research (ATOMKI), Debrecen, Hungary}}

\newcommand{\halves}[1]{{#1}/{2}}

\DeclareMathOperator{\sign}{sgn}
\begin{document}

\title{The $^{3}$He($\alpha,\gamma$)$^{7}$Be $\gamma$-ray angular distribution: measured, and calculated by effective field theory}

\author{S.~Turkat} \tu 

\author{P.~Hempel}\hzdr\tu

\author{X.~Zhang}\email{zhangx@frib.msu.edu}\michigan

\author{S.~Akhmadaliev}\hzdr

\author{D.~Bemmerer}\email{d.bemmerer@hzdr.de}\hzdr

\author{A.~Boeltzig}\hzdr

\author{A.~Caciolli}\padova \INFN

\author{A.~Freimann}\hzdr\tu

\author{M.~Grieger}\hzdr

\author{S.~Hammer}\hzdr

\author{T.~Hensel}\tu\hzdr

\author{L.~Hübinger}\hzdr

\author{N.~A.~Hofmann}\hzdr

\author{J.~Koch}\tu

\author{T.~Lossin}\hzdr

\author{F.~Ludwig}\hzdr\tu

\author{E.~Masha}\hzdr

\author{K.~M.~Nollett}\sandiego

\author{M.~Osswald}\tu

\author{S.~Reinicke}\hzdr\tu

\author{S.~Rümmler}\hzdr\tu

\author{K.~Schmidt}\hzdr

\author{K.~Stöckel}\hzdr\tu

\author{T.~Szücs}\atomki\hzdr

\author{S.~Vincent}\hzdr\tu

\author{L.~Wagner}\tu\hzdr

\author{J.~Wünsche}\tu\hzdr

\author{A.~Yadav}\hzdr\tu

\author{K.~Zuber}\tu

\date{\today}

\begin{abstract}

The $^{3}$He($\alpha,\gamma$)$^{7}$Be reaction takes place in several distinct astrophysical scenarios. First, in solar hydrogen burning, $^{3}$He($\alpha,\gamma$)$^{7}$Be controls the competition between the pp-I chain, on the one hand, and the pp-II and -III chains, on the other hand. Its rate directly affects the solar model predictions for the well-measured $^7$Be and $^8$B solar neutrino fluxes, which can in principle be used as solar thermometer. Second, in Big Bang nucleosynthesis, $^{3}$He($\alpha,\gamma$)$^{7}$Be is the main production channel for primordial $^7$Li, the decay product of $^{7}$Be. Finally, the reaction is believed to produce radioactive $^7$Be in astrophysical novae. The cross section of this reaction has been studied previously using different methods, namely $^{7}$Be activation, $^{7}$Be recoil detection, and in-beam $\gamma$-ray detection. The latter studies have in many cases been performed with just one detector in close geometry. Here, for the first time a comprehensive measurement of the $^{3}$He($\alpha,\gamma$)$^{7}$Be $\gamma$-ray angular distribution is reported. Altogether 22 high-purity germanium detectors were used, arranged at 27-145$^\circ$ angle with respect to the beam direction. Data were taken at 442-1213\,keV center-of-mass energy. In addition, one data point for the astrophysical S-factor was obtained with the activation method.
The angular data are compared with the predictions from a hybrid theoretical approach, where the E1 transitions are modeled using a cluster-based effective field theory (EFT) while the M1 and E2 transitions within an EFT-inspired framework. Good agreement is found, if the interference among these multipolarities is taken into account in the theory.

\end{abstract}

\maketitle


In the center of the Sun (temperature $T\approx16$\,MK), the $^{3}$He($\alpha,\gamma$)$^{7}$Be reaction  (shorthand $S_{34}$) controls the passage from the proton-proton I (pp-I) to the pp-II and pp-III chains \cite{Orebi21-ARNPS}, hence the fluxes of $^7$Be and $^8$B neutrinos. The $^7$Be and $^8$B neutrino fluxes are now known from observations with 1.8\% and 1.9\% precision, respectively, assuming the neutrino flavour oscillation parameters to be well understood \cite{2024JHEP...02..064G}. The $^7$Be and $^8$B neutrino fluxes depend almost linearly on the $S_{34}$ rate, with a partial derivative of 0.853 and 0.806, respectively \cite{2013PhRvD..87d3001S}. In contrast,  $S_{34}$ is currently affected by 5.1\% uncertainty, according to the recent Solar Fusion III (SF-III) decadal review \cite{2025RvMP...97c5002A}.

In order to improve this situation, the SF-III community review makes two recommendations: first, a cross section measurement covering a wide energy range, and second, a new study of the $\gamma$-ray angular distribution \cite{2025RvMP...97c5002A}. The present work addresses the latter recommendation, and facilitates the extrapolation from $E \sim$ 1 MeV, where several multipolarities play a role \cite{Zhang20-JPG}, down to solar energies, where E1 dominates. 

$^{3}$He($\alpha,\gamma$)$^{7}$Be also takes place in two further astrophysical environments beyond the Sun. In the first scenario, Big Bang nucleosynthesis (BBN), primordial $^7$Li production is believed to be dominated by $^{3}$He($\alpha,\gamma$)$^{7}$Be \cite{2024PhRvD.110c0001N} and (much later) subsequent electron capture decay. Cosmological parameters inferred from the cosmic microwave background \cite{Planck18_CosmologicalParameters} imply about 2-4 times as much $^7$Li as the amount found in ancient stars; this is the cosmic lithium problem \cite{Fields11-ARNPS}. Direct $^{3}$He($\alpha,\gamma$)$^{7}$Be data over much of the BBN range \cite[e.g.]{Bemmerer06-PRL,Confortola07-PRC,Brown07-PRC,DiLeva09-PRL,Kontos13-PRC} and also other nuclear reaction studies \cite[e.g.]{2018PhRvL.121d2701D,2022PhRvL.128y2701A,2024PhLB..85338673A} have ruled out a nuclear solution to this problem. Non-nuclear solutions inside the stars where lithium is observed \cite{2024A&A...690A.245B} or by exotic particle decays \cite{Sato2022} in the early universe can only be assessed with precise $^{3}$He($\alpha,\gamma$)$^{7}$Be data as a baseline. The final astrophysical scenario relevant for $^{3}$He($\alpha,\gamma$)$^{7}$Be are classical novae, which are known to produce $^7$Be \cite{2016MNRAS.463L.117M} and believed to be at the origin of much of the galactic $^7$Li \cite{2024ApJ...962..191S}.

\begin{figure}[tb]
\includegraphics[width=\columnwidth,trim=5mm 7mm 13mm 0mm]{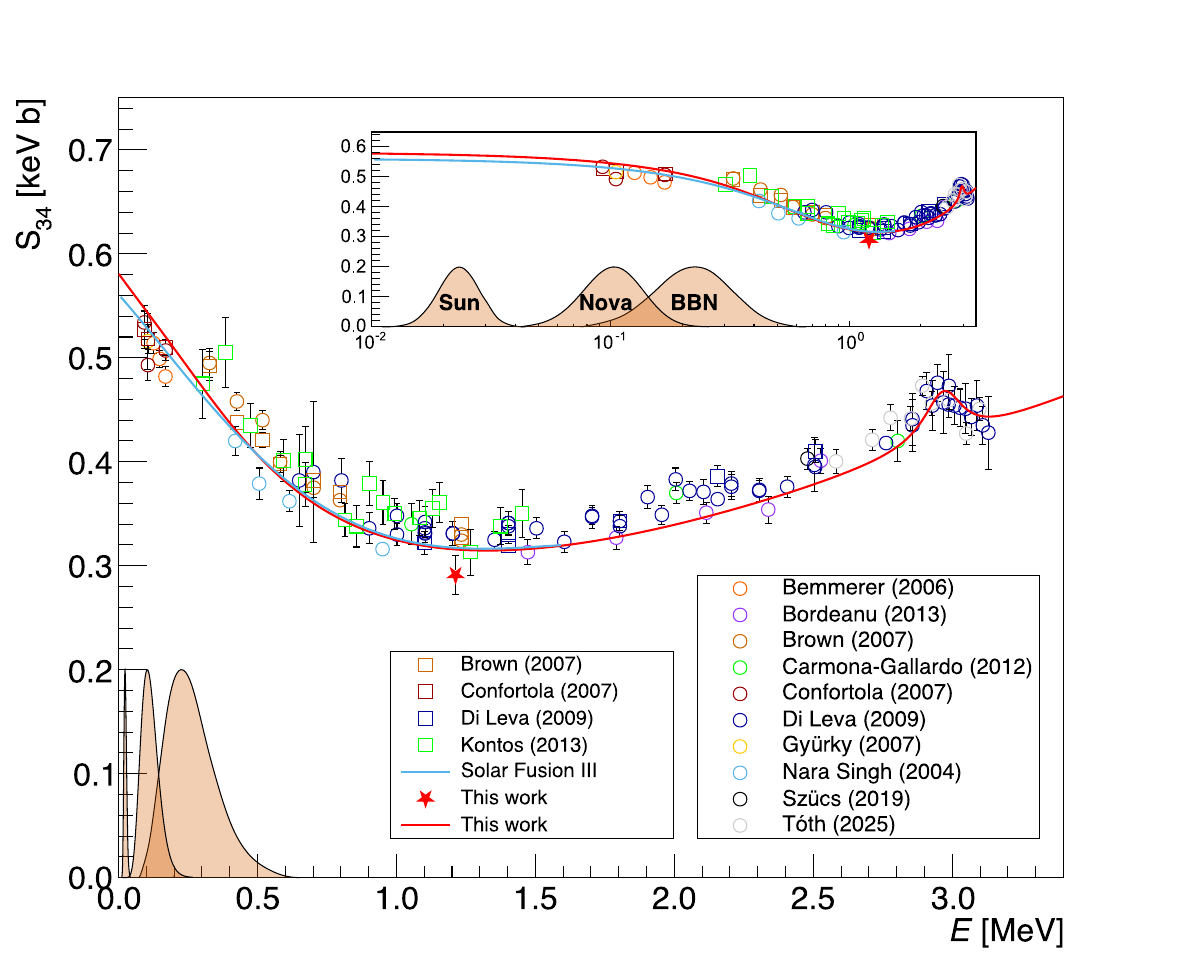}
\caption{\label{fig:SFactor} $^3$He($\alpha$,$\gamma$)$^7$Be data for the astrophysical S-factor $S_{34}$. Prompt-$\gamma$ data are shown as squares \cite{Brown07-PRC, Confortola07-PRC, Kontos13-PRC,DiLeva09-PRL}, activation and recoil data as circles \cite[and present work]{2004PhRvL..93z2503S,Bemmerer06-PRL, Gyurky07-PRC, Confortola07-PRC, Brown07-PRC, DiLeva09-PRL, Carmona12-PRC, Bordeanu13-NPA, 2019PhRvC..99e5804S}.  The SF-III fit \cite{2025RvMP...97c5002A} is shown in blue, the present fit in red. Calculated sensitivity curves for the Sun and for BBN \cite{Takacs15-PRD} as well as for novae are shown in orange.}
\end{figure}

The $^{3}$He($\alpha,\gamma$)$^{7}$Be reaction has been studied previously using the activation \cite{2004PhRvL..93z2503S,Bemmerer06-PRL, Gyurky07-PRC, Confortola07-PRC, Brown07-PRC, DiLeva09-PRL, Carmona12-PRC, Bordeanu13-NPA, 2019PhRvC..99e5804S,2023PhRvC.108b5802T,2025EPJA...61..274T}, recoil detection \cite{DiLeva09-PRL}, and prompt-$\gamma$  \cite{Brown07-PRC, Confortola07-PRC, Kontos13-PRC,DiLeva09-PRL} methods, with a few works using several of these techniques (Figure \ref{fig:SFactor}). Recent works by the prompt-$\gamma$ technique used just one detector at 90$^\circ$ angle to the beam direction \cite{Brown07-PRC, Kontos13-PRC}, making the data dependent on a correction for the $\gamma$-ray angular distribution from a reaction model \cite{1963PhRv..131.2582T}. One of these groups also conducted activation measurements \cite{Brown07-PRC}, and they found slightly better agreement between prompt and activation data when omitting the correction. Given this situation and the fact that the $^{3}$He($\alpha,\gamma$)$^{7}$Be $\gamma$-ray angular distribution has hitherto only been measured using just three angles \cite{1982ZPhyA.304..307K}, there have been calls for new angular data \cite{Adelberger11-RMP,Kontos13-PRC} that have recently been repeated in SF-III \cite{2025RvMP...97c5002A}. In addition, theoretical work suggests a correlation between the $\gamma$-ray angular distribution at high energy, $\sim$1 MeV, and the extrapolated $S_{34}$ factor at solar energy \cite{Zhang20-JPG}, further underlining the need for new data.


In order to address these points, a new measurement of the $^{3}$He($\alpha,\gamma$)$^{7}$Be $\gamma$-ray angular distribution is reported here. It was divided in two campaigns called A and B, respectively. Campaign A took place at the Felsenkeller shallow-underground 5\,MV Pelletron accelerator \cite{2025EPJA...61...19B} in Dresden, Germany. The $^4$He$^+$ ion beam of $E_\alpha$=1101-2881\,keV passed a 90$^\circ$ bending magnet, a water-cooled copper collimator (10\,mm diameter), and a liquid nitrogen (LN$_2$) cooled copper pipe (300 mm long, 18\,mm inner diameter) and then hit the target (also LN$_2$-cooled to limit out-diffusion of $^3$He) with typically 10-20\,$\upmu$A current. Secondary electrons were suppressed by up to -1500\,V bias applied to the copper pipe and by additional permanent magnets. During the irradiation, the target was visually monitored with an infrared-sensitive camera. A thermoresistor at the back of the target showed a typical temperature of -120\,$^\circ$C to -140\,$^\circ$C depending on the target current.

The targets were tantalum disks (0.22\,mm thick, 27\,mm diameter), hereafter called ST9-ST11 and ST15-ST17. They were pre-implanted with a fluence of $10\times10^{17}$ cm$^{-2}$ $^{3}$He at the HZDR Ion Beam Center (IBC) at $E(^3\text{He})=35$\,keV, with a homogeneously wobbled beam and LN$_2$ cooling.
Additional, non-implanted disks called Ta51–Ta55 and Ta61 were used to study the beam-induced background. 

The $\gamma$ rays were detected by two Euroball \cite{1997ZPhyA.358..139S} septuple clusters (hereafter called EB17, EB18), two Miniball \cite{2001PrPNP..46..389E} triple clusters (MB1, MB2), and two single crystal HPGe detectors (Can60, Ron100). EB17, MB1, MB2, Can60, and Ron100 were positioned horizontally at angles of 37$^\circ$–145$^\circ$ and 15-20\,cm distance around the target.  EB18 was positioned above the target at 90$^\circ$ at a distance of 5\,cm. 
The setup was modeled in GEANT4 \cite{2003NIMPA.506..250A}. The model was validated using radionuclide standards, leading to a final $\gamma$-ray efficiency uncertainty of 2.2\%. 

During the $^4$He beam experiment, the targets were monitored using the differential $\gamma$-ray yield. After an initial decline of 30\%, the target thickness in the irradiated area was found to be stable within 10\%, with the exception of target ST11 ($E$ = 555 keV) which degraded linearly by 80\%. Visual inspection after the runs showed no evidence of $^4$He blistering.

Campaign B preceded campaign A and produced just one data point. It was conducted at the 3\,MV Tandetron accelerator at IBC, with a $^4$He$^{2+}$ beam of 6-8\,$\upmu$A (electrical) and $E_\alpha=2882$\,keV. For the {\it in situ} target characterization, a $^2$H$^+$ beam of 0.1\,$\upmu$A was used. The targets called Ta100 and Ta502 were each pre-implanted with $^3$He at $E(^3\text{He})=40$\,keV (10 and 5$\times10^{17}$ cm$^{-2}$, respectively) and $10$\,keV (2.4 and 1.2$\times10^{17}$ cm$^{-2}$, respectively).  For target Ta502, nuclear reaction analysis showed that $\sim$75\% of the implanted $^3$He was found in the target at the beginning of the main experiment. The emitted $\gamma$ rays were detected in four HPGe detectors at 45$^{\circ}$, 90$^{\circ}$, 135$^{\circ}$, and 153$^{\circ}$.

\begin{figure*}[tb]
\includegraphics[width=\textwidth]{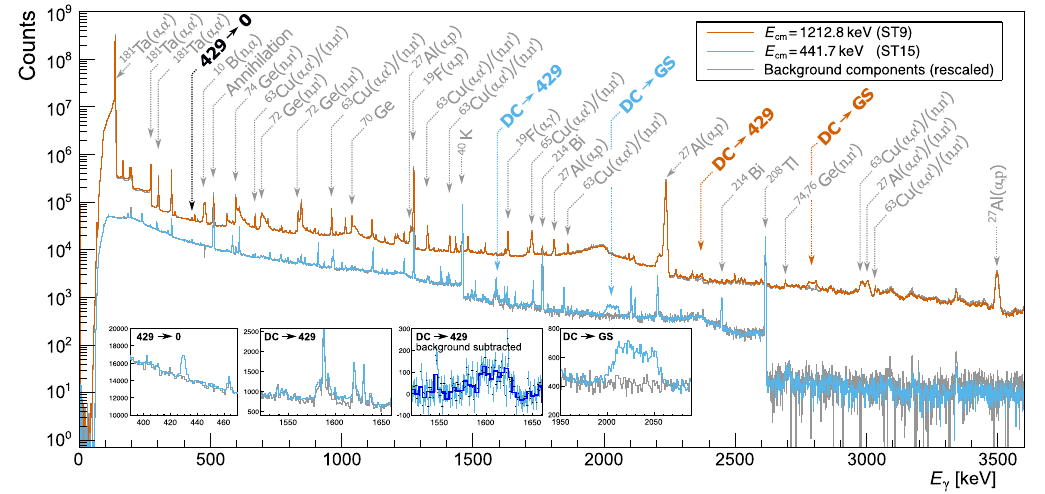}
\caption{\label{fig:InBeamSpectra} Pulse height spectra at 90$^\circ$ (sum of EB18 subdetectors B, G, and E) for the lowest (blue, $E$ = 441.7 keV) and the highest (orange, 1212.8 keV) irradiation energies, respectively, campaign A. The background (see text) is shown in grey. The insets are for $E$ = 441.7 keV. The DC$\rightarrow$429 region includes $^{232}$Th chain background lines at 1588, 1621, and 1631 keV.}
\end{figure*}


The in-beam experimental spectra (Figure \ref{fig:InBeamSpectra}) clearly show a box-like, 40-50 keV wide $\gamma$ peak for ground state capture, reflecting the target profile. The background (shown in grey) is estimated in two steps: for the beam-induced background, a blank target is irradiated at the same $\alpha$ beam energy. For the room background, runs taken during beam-off days were used. The only significant differences between the spectra with the $^3$He targets and the background spectra are the $\gamma$ rays from the reaction under study.

\begin{figure*}[tb]
\includegraphics[width=\textwidth,trim=0 3mm 0 5mm,clip]{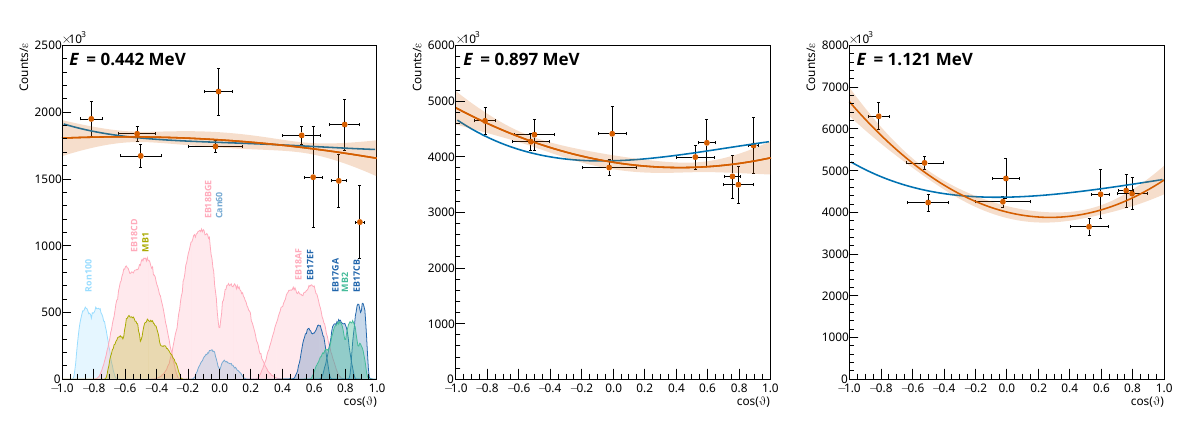}
\caption{\label{fig:AngularDistribution} $\gamma$-ray angular distributions at $E$ = 0.442, 0.897, and 1.121 MeV (left to right).  In the left panel, the detection efficiency as a function of angle per detector is shown in arbitrary units. The curves are the present Bayesian fits (red) and the previous theory by Tombrello \cite[blue]{1963PhRv..131.2582T}.}
\end{figure*}

For each experimental run, an angular distribution fit was performed (Figure \ref{fig:AngularDistribution}). For the fit, the full-energy peak detector responses $\eta_i(\cos \vartheta)$, where $\vartheta$ is the angle between the directions of the beam and of the emitted photon, were modeled taking into account the precise detector geometry, Doppler and recoil corrections, in GEANT4. Then, the experimental net counts $N_i$ ($i \in \{ 1,..., 22 \}$) for each of the 22 detectors were fitted with the Bayesian Analysis Toolkit (BAT) \cite{Caldwell_2009}. The corresponding log-likelihood function 
was optimised by adjusting the coefficients $a_{0,1,2}^j$ ($j \in \{\rm GS, ES\}$) for the Legendre polynomials of orders 0-2. 

The Bayesian fit includes two different nuisance parameters for the systematic uncertainties: first, the $\gamma$-ray detection efficiency normalization (2.2\%), and second, the uncertainty of the modeled angular dependent efficiency (Figure \ref{fig:AngularDistribution}). In order to derive their effects on the fit results, the efficiencies of the individual detectors were randomly varied and the model re-fitted many times, resulting in $\Delta a_1^{\rm GS,syst}$ = 0.002 and $\Delta a_2^{\rm GS,syst}$ = 0.003. As a cross-check, the effect on the fitted normalization (expressed by the 0 order Legendre coefficient) was found to be $\Delta a_0^{\rm GS,syst}/a_0^{\rm GS}$ = 2.2\%, as expected.

The data for the ground state transition show isotropy, $a_1 \approx a_2 \approx 0$, at the lowest energy explored, $E$ = 441.7 keV, and a slowly increasing anisotropy towards 1212.8 keV (Figure \ref{fig:Legendre_GS}). 
The model (see below) describes the increase in $a_2$ and the decrease in $a_1$ with rising energy well up to an energy of $E$ = 900 keV. The highest energy data point at 1212.8 keV shows higher $a_2$ and lower $a_1$ than the model. The data for the transition to the excited state (not shown) are less precise due to higher background but generally agree with the model. 

From the fitted $a_0$ coefficients for the excited and ground state transitions, the experimental branching ratios were determined. They were found to be independent of energy within uncertainties, with an average ratio of $\gamma_1/\gamma_0 = a_{0}^\textrm{ES}/a_{0}^\textrm{GS}$ = 0.34$\pm$0.02. Using the isotropic 429\,keV $\gamma$ ray in the 90$^\circ$ detectors, a consistent branching ratio of $\gamma_{429}/\gamma_0$ = 0.36$\pm$0.02 is found. 

These values can be compared with literature branchings obtained at one angle: $\gamma_{429}/\gamma_0$ = 0.426$\pm$0.003 \cite{Brown07-PRC} and 0.39$\pm$0.01 \cite{Kontos13-PRC}. When taking the present average $a_2$ value for $\gamma_0$ into account, these literature values have to be corrected downward to 0.402$\pm$0.003 and 0.37$\pm$0.01, respectively, closer to the present $\gamma_1/\gamma_0$. 


In campaign B, in addition to the angular study, also the absolute astrophysical S-factor was measured at one energy. 
The areal $^3$He density of the target was determined relative to the $^3$He($d,p$)$^4$He cross section which is known to 4.3\% \cite{Wielunska16}. Using a silicon surface barrier detector with an active area of 300\,mm$^2$ and a sensitive thickness of 2\,mm at an angle of $\vartheta=144.5^\circ$, a $^3$He thickness in Ta of (4.65$\pm0.12^{\rm stat}\pm0.20^{\rm syst})\times10^{17}$ cm$^{-2}$ was found. The activation analysis at the Felsenkeller TU1 detector \cite{Turkat23-APP} gave a $^7$Be activity of 115$\pm7^{\rm stat}\pm2^{\rm syst}$ mBq. 
The resulting astrophysical $S$ factor is $S_{34}$(1212.9 keV) = 0.291 $\pm$ 0.019$^\textrm{stat}$ $\pm$ 0.014$^\textrm{syst}$ keV\,b (Figure \ref{fig:SFactor}), with the error dominated by the target thickness uncertainty.


We turn now to the theoretical treatment. Three different multipoles are considered: E1, E2, and M1 (Table~\ref{tab:multipoles_channels}). One of them,  E2, is affected by the $7/2^-$ resonance located at $E_{\rm cm} \approx$ 3 MeV (Figure \ref{fig:SFactor}, Table~\ref{tab:multipoles_channels}). 

Taking into account the angular momentum barrier, it is expected that E1 dominates the total $S_{34}$ factor, on the order of 0.3-0.5 keV\,b, and that the M1 and E2 contributions are smaller, on the order of $10^{-3}$ keV b, up to 2 MeV energy \cite{Nollett:2001ub}. However, for the $\gamma$-ray anisotropy, 
interferences between E1 and the other two (M1 and E2) could be significant, which give rise to non-zero first order Legendre coefficients $a_{1}$ for the ground and excited states---but not to the second order Legendre coefficient $a_{2}$. Below the resonance, $a_2$ is dominated by the E1 component alone, but near the resonance region, the interferences among M1 and E2 could be important.

To model the E1 amplitudes and the related $a_2$, we employ the next-to-leading order (NLO) cluster-based effective field theory (EFT) framework developed in Ref.~\cite{Zhang20-JPG}. Importantly, the EFT treats $^3$He and $\alpha$ as basic degrees of freedom (DOF) and $^7$Be as a composite of the two. In order to calibrate E1 via Bayesian analysis, as in SF-III, the experimental cross section data are fitted, adding now as additional constraints the present angular distribution data and the present branching ratio, with $E$ up to 2 MeV. We use the same flat prior distributions for the parameters of this NLO EFT as those used for in SF-III~\cite{2025RvMP...97c5002A}.
The resulting total $S_{34}$ factor (Figure \ref{fig:SFactor}) is overall very close to the Higa {\it et al.} \cite{2018EPJA...54...89H} based fit adopted in SF-III \cite{2025RvMP...97c5002A}. However, at low energy, relevant to the Sun, the present $S_{34}$ is higher by 0.02 keV\,b, still within the SF-III error bar. In Figure~\ref{fig:SFactor}, we also extrapolate the E1 component of the $S$ factor and include the other multipoles (discussed below) for the energies above 2 MeV. The resulting $a_2$ for the ground state will be addressed together with the $a_1$ after we introduce M1 and E2.  

\begin{figure}[tb]
\includegraphics[width=\columnwidth, trim = 10mm 0mm 18mm 0mm]{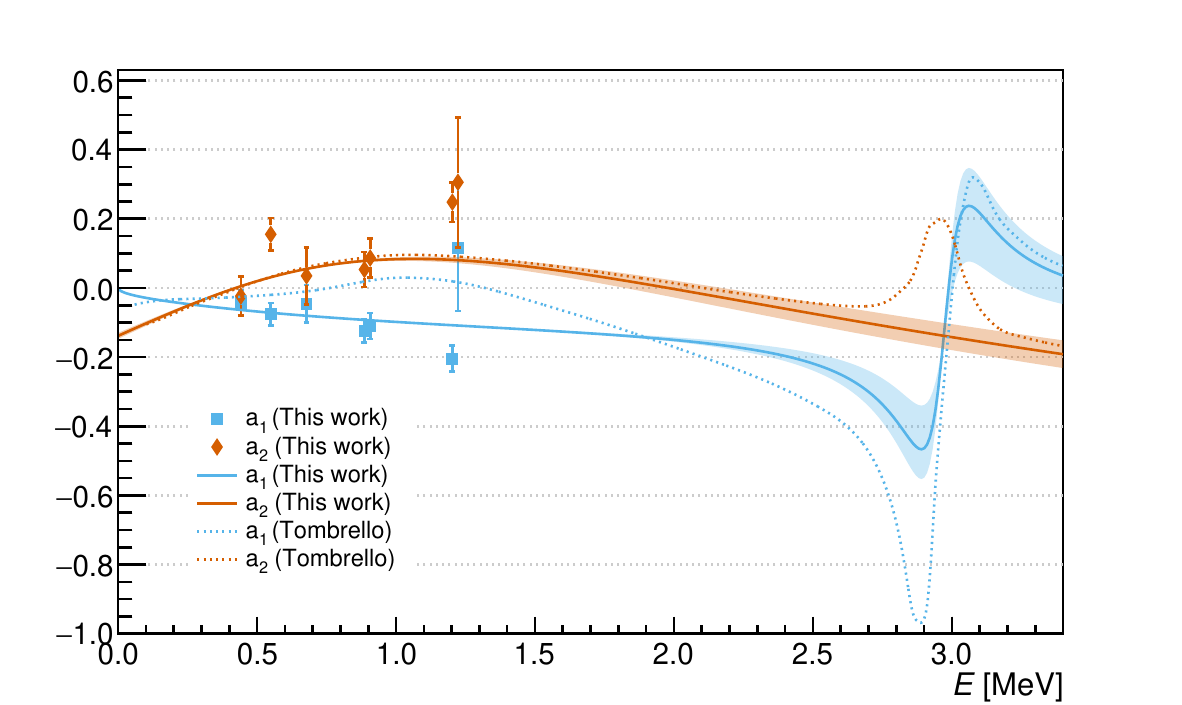}
\caption{\label{fig:Legendre_GS} Legendre coefficients $a_{1,2}^{\rm GS}$ as a function of $E$: experiment (with statistical error bars) and theory (straight line with shaded error band). The previous curves by Tombrello \cite{1963PhRv..131.2582T} are shown as dashed lines. See text for details.}
\end{figure}

To model the M1 and E2 multipoles, we implement a phenomenological approach inspired by previous cluster-EFT studies of capture reactions~\cite{Zhang20-JPG,Zhang:2017yqc,Zhang:2015ajn,Zhang:2014zsa,Zhang:2013kja}. We, again, consider $^3$He and $\alpha$ as the basic DOF. Moreover, the EFT-based E1 results are extrapolated to the resonance region when considering the interference between the E2 and M1 multipoles and the E1.  
This interference could be important even when total contributions are small.  The M1 transitions from $p$-wave initial states and E2 transitions from $f$-wave may be written in terms of their characteristic dependencies on photon energy $\omega$ and on overlap integrals of the radial wave function
($r$ for the relative radial coordinate between $^3$He and $\alpha$): 
\begin{align}
  \langle \ell=1, j' || \hat{T}^\mathrm{ma}_{1} || 1,j \rangle  & \propto  \frac{\omega}{2 M_N}  \bigg\{  \int_{r_c}^\infty dr\,r^2\, C_{j'} \phi_{1,j'}(r)\, \psi_{1,j}(E,r)  \notag \\ 
   & \hspace{1.5cm} +  L_{j'j}(r_c) f_{j'j}(E) \bigg\}\,,
	\label{eq:T1ma}
    \\ 
   \langle 1, {j'} || \hat{T}^\mathrm{el}_{2} || 3, j \rangle  & \propto  \omega^2\Bigg\{ \int_0^\infty dr\,r^2  \, C_{j'} \phi_{1,j'}(r)\, r^2 \phi_{3,j}(E,r) \notag \\ 
  +  \delta_{j,\halves{7}} & f(E)
       \frac{ \sign(B(E2)) \sqrt{B(E2)  \Gamma(E)} }{E - E_R + i  \Gamma(E)/2 }  \Bigg\}  \,.
	\label{eq:T2el}
\end{align}
Here, $\hat{T}^\mathrm{ma}_{1}$ and $\hat{T}^\mathrm{el}_{2}$ are the M1 and E2 multipole operators; the initial and final states in the reduced matrix elements are labeled by the orbital angular momentum $\ell$ and the respective total angular momentum $j$ and $j'$. In the right sides of these expressions, the wave functions $\phi(E,r)$ and $\psi(E,r)$ are indexed by their inter-cluster orbital and total angular momentum. Note that all the involved incoming channels (including those for the E1, see Table~\ref{tab:multipoles_channels}) are decoupled due to their either different total angular momenta or parities. Moreover, in these equations, $M_N$ is the nucleon mass (present in the definition of nuclear magneton), $E$ is the relative $^3$He-$\alpha$ energy, $C$s the asymptotic normalization coefficients (ANCs) related to $\phi$ and $\psi$, and $E_R$, $\Gamma(E)$ and $B(E_2)$ the energy location, the energy-dependent decay width and  the BE(2) transition strength of the 3 MeV resonance. These parameters and the other symbols, $L_{j'j}(r)$, $f_{j'j}(E)$ and $f(E)$ will be discussed further below.

\begin{table}[t]
    \begin{ruledtabular} 
    \begin{tabular}{ c c c c} 
      State     &  E1 & M1 & E2  \\ \hline
      GS ${{3}/{2}}^-$ &  $S_{1/2}$, $D_{5/2,3/2}$  & $P_{3/2,1/2}$  & 
 $P_{3/2,1/2}$, $F_{7/2,5/2}$    \\ \hline 
      ES ${{1}/{2}}^-$ & $S_{1/2}$, $D_{3/2}$  & $P_{3/2,1/2}$   & $P_{3/2}$, $F_{5/2}$   \\ 
    \end{tabular} 
    \caption{\label{tab:multipoles_channels} Multipoles and their initial states considered here.     }
     \end{ruledtabular}
\end{table}

For M1, in Equation~(\ref{eq:T1ma}), $C_{j'}\phi_{1,{j'}}(r)$ is the asymptotic wave function (between $^3$He and $\alpha$) with $C_{j'}$ the corresponding ANC of the ground or excited ($p$-wave) states of $^7$Be. Meanwhile, $\psi_{1,j}(E,r)$ is the asymptotic form of the incoming inter-cluster $p$-wave scattering wave function, which depends on the Coulomb and the strong interaction phase shift. 

We model the $p$-wave phase shifts by a $T$-matrix consisting only of below-threshold poles corresponding to the bound states. 
As a result, the incoming wave functions are determined by the bound state properties, including ANCs and binding energies.

Since the M1 radial overlaps use the asymptotic wave functions, the integral is divergent. We apply an integral cut-off at small radius $r_c$ to regularize the integral, and more importantly, include the $L_{j'j}(r_c)$ parameter to correct the regularization-induced errors, as inspired by the EFT approach~\cite{Zhang20-JPG}. Also note that the $f_{j'j}(E)$ factor, whose exact form can be found in Ref.~\cite{Zhang20-JPG}, has a nontrivial energy dependence induced by the corresponding strong interaction phase shift and Gamow suppression factor. Therefore, this approach is different from the previous external capture calculations \cite[e.g.]{1963PhRv..131.2582T} that set $r_c$ to a physics-informed value without any corrections. 
 
Specifically, $r_c$ = 6 fm is chosen, as informed by  microscopic calculations \cite{Nollett:2001ub,2011PhRvL.106d2502N}, and $L_{j'j}$ (three of them) are fixed according to the $^7$Be ground and excited states' magnetic moments, namely $\mu_{gs} = -1.398$ \cite{TILLEY20023} and $\mu_{es} = 1.08$\footnote{From our own variational Monte Carlo calculation.}, and the transition reduced matrix element to 2.74 magnetons \cite{Pastore:2012rp}. To that end, we use Eq.~\eqref{eq:T1ma} with $\psi_{1,j}$ substituted by the bound state wave function $C_{j}\phi_{1,{j}}(r)$ to relate $L_{j'j}$ to the corresponding $M1$ transition or the magnetic moment. In short, we have a complete theory for M1 after fixing $L_{j'j}$, which predicts the E1--M1 interference and its contribution to the $a_1$.

For E2, Eq.~(\ref{eq:T2el}) shows the amplitudes for the inter-cluster $f$-wave initial state. The first term $\phi_{3,j}(E,r)$ is the regular Coulomb wave function with angular momentum $3$. The resonance term led by the Kronecker $\delta_{j,7/2}$ approximates the corresponding $T$-matrix by the resonance pole contribution. The resonance width $\Gamma(E)$ \cite{TILLEY20023} with its energy dependence as a $f$-wave resonance is then used, as well as the reduced electric quadrupole transition probability $B(E2) = 52\pm 31$ $e^2\mathrm{fm}^4$ value extracted\footnote{In a separate analysis, we subtract the E1 component of the $S$ factor (via extrapolating the mean value of our $E1$ prediction to the resonance region) from the total $S$ factor data and then fit the $B(E2)$ to the residual around the resonance. The resulting $B(E2)$ is consistent with those of Ref.~\cite{DiLeva09-PRL}.} from the measurement of $S_{34}$ near the 3\,MeV resonance ~\cite{DiLeva09-PRL}. Note that $f(E)$ is a parameter-free $E$-dependent function composed of Gamow and kinematic factors. For the other $f$-wave channels, no resonance exists and thus the incoming wave function is approximated by the Coulomb wave function. 
Similarly, for the $p$-wave initial states in the E2 transitions, the $\phi_{3,j}(E,r)$ in Eq.~\eqref{eq:T2el} is substituted by $\psi_{1,j}(E,r)$ as used in the M1 amplitude. Here, no $r_c$ cut-off is needed as the E2 integrals converge, and no extra correction terms are included.

By combining the E1 results with the modeling of M1 and E2, we can compute the total $S$ factor even near the resonance region where the E2 is significant. We can also calculate $a_{1}^{\rm GS,\,ES}$ coefficients due to the interferences between E1 and the other two multipoles, as additions to the aforementioned $a_{2}^{\rm GS,\,ES}$.

Now, we compare measured and calculated $a_{1,2}^{\rm GS}$ coefficients (Figure \ref{fig:Legendre_GS} for the ground state). The error band of $a_{2}^{\rm GS}$ comes from the Bayesian fit of the EFT-based E1 multipole. The $a_{1}^{\rm GS}$ error band is due to the variation of $B(E2)$ of the 3\,MeV resonance. While the general trend of the data is well reproduced, the last data point at $E \approx$ 1.2 MeV shows higher absolute values for both $a_{1}^{\rm GS}$ and $a_{2}^{\rm GS}$. Further data at higher energy are needed to determine whether this is just an outlier or a real effect. Overall, the good agreement between data and $a_{1}^{\rm GS}$ shows that the interference effects between E1, E2, and M1 are correctly modeled here. Still, a dedicated measurement of the anisotropy near the 3\,MeV resonance with its strong predicted interference pattern would add additional insight on this point. Note the lack of resonance behavior of the EFT-based $a_2^{\rm GS}$ is expected, as we only consider the E1 multipole, which is not coupled to the resonance. 

Considering the impact of the present data and calculations on the literature, Brown \cite{Brown:2007sj} and Kontos \cite{Kontos13-PRC} used one detector at 90$^\circ$, necessitating a correction for non-zero $a_2^{\rm GS}$. They used Tombrello \cite{1963PhRv..131.2582T}, which is very close to the present $a_2^{\rm GS}$ (Figure \ref{fig:Legendre_GS}). The $\gamma$-ray part of the LUNA study used one detector at 55$^\circ$ \cite{Confortola07-PRC} and adopted Tombrello \cite{1963PhRv..131.2582T}. The present $a_1^{\rm GS}$ are close to Tombrello at the low energies of LUNA, $E_{\rm}$ = 0.09-0.17\,MeV, but not at 0.5-1.5 MeV (Figure \ref{fig:Legendre_GS}). 
Finally, ERNA used three large detectors covering much of a hemisphere \cite{2008NIMPA.595..381D,DiLeva09-PRL}, mitigating any angular effects. 

Finally, the present fits can be used to study the correlation between angular coefficients at high energy and zero-energy extrapolated $S_{34}(0)$. At $E$ = 1.2 MeV, near the predicted value of $a_2^{\rm GS}$(1.2\,MeV) = 0.08, a slope of $\partial S_{34}(0) / \partial a_2^{\rm GS}(1.2\,{\rm MeV})$ = -1.9 keV\,b is found. The physical reason for this correlation is that at low energies the E1 component that produces the second order Legendre anisotropy strongly dominates, so that weak angular effects near 1.2 MeV may be used to probe the low-energy S-factor. In order to match $\Delta S_{34}(0)$ = 0.028 keV\,b \cite[SF-III]{2025RvMP...97c5002A}, a precision of $\Delta a_2^{\rm GS}$(1.2\,MeV) $\sim$ 0.01 would be required. With negligible background, like in a jet gas target \cite{2026NIMPA108271034Y}, this would be achievable with $\sim 10^4$ detected counts per detector. 

Summarizing, the previously recommended \cite{Adelberger11-RMP,Kontos13-PRC,2025RvMP...97c5002A} experimental study of the $\gamma$-ray angular distribution of the $^3$He($\alpha,\gamma$)$^7$Be reaction has been carried out. The new angular data are matched by effective field theory based calculations. A forward-backward anisotropy described by a first-order Legendre coefficient $a_1$ has been found and explained by interferences between the weak E2 and M1 multipoles and the dominating E1. The computations show a correlation between the $a_2$ angular coefficient at 1.2\,MeV and the zero-energy S-factor $S_{34}(0)$. 
The branching ratio for capture to the first excited state of $^7$Be has been remeasured and was found to be somewhat lower than in previous work, and an angular correction for previous branching ratios has been suggested.  One new activation-based S factor data point is presented at 1212.9 keV. 

Concluding, the previously unknown $^3$He($\alpha,\gamma$)$^7$Be $\gamma$-ray angular distribution has been measured, and modeled in the EFT framework including interference effects, removing a significant source of uncertainty \cite{2025RvMP...97c5002A} on this crucial solar fusion and Big Bang reaction.

\begin{acknowledgments}

The authors are indebted to Bernd Scheumann and Claudia Neisser for sample preparation, and to Andreas Hartmann for technical support. 
--- 
The use of two septuple detectors (EB17, EB18) from the GAMMAPOOL and financial support by the European Union (ChETEC-INFRA 101008324 and LUNANOVA 101224374), Deutsche Forschungsgemeinschaft DFG (ZU123/21-1, BE4100/4-1), the Helmholtz Association (ERC-RA-0016 and VH-NG-22-08), Konrad-Adenauer-Stiftung, and the Hungarian National Research, Development, and Innovation Office (Grant No. NKKP ADVANCED 152233) is gratefully acknowledged. X.Z. acknowledges support from the U.S. Department of Energy, Office of Science, Office of Nuclear Physics, under the FRIB Theory Alliance award DE-SC0013617.

\end{acknowledgments}



\end{document}